\documentclass[twocolumn,twoside,slac_two]{revtex4}
\usepackage{fancyhdr}
\def\Journal#1#2#3#4{{#1} {\bf #2}, #4 (#3)} 

\def\NIM{Nucl.\,Instrum.\,Methods}

\def\NP{Nucl.\,Phys.}
\def\NPBP{Nucl.\,Phys. B (Proc. Suppl.)}

\def\PL{Phys.\,Lett.}

\def\JPG{J.\,Phys. G}
\def\PRL{Phys.\,Rev.\,Lett.}

\def\PR{Phys.\,Rev.}

\def\JP{J.\,Phys.}
\def\AP{Astropart.\,Phys.}

\def\ARNPS{Annu.\,Rev.\,Nucl.\,Part.\,Sci.}

\def\bbn{$\beta\beta$-0$\nu$}
\def\bb{$\beta\beta$-2$\nu$}
\def\Qbb{$Q_{\beta\beta}$}
\def\taubb{$\tau_{1/2}^{0\nu}$}
\def\meff{$\langle m_\nu\rangle$}
\def\cky{c/kg/y}
\def\ckky{c/keV/kg/y}
\def\ckty{c/keV/t/y}
\def\ky{kg$\times$y}
\def\my{moles$\times$y}
\def\Te{$^{130}$Te}
\def\Ge{$^{76}$Ge}
\def\Ca{$^{48}$Ca}
\def\Cd{$^{116}$Cd}
\def\Nd{$^{150}$Nd}
\def\Gd{$^{160}$Gd}
\def\Se{$^{82}$Se}
\def\Mo{$^{100}$Mo}
\def\Xe{$^{136}$Xe}

\def\Co{$^{60}$Co}
\def\Tl{$^{208}$Tl}
\def\teo{TeO$_2$}

\def\me{$m_e$}

\begin{document}

\title{Double beta Decay: Experiments and Theory Review}

\author{A. Nucciotti\footnote{E-mail: angelo.nucciotti@mib.infn.it}}
\affiliation{Dipartimento di Fisica "G. Occhialini", Universit\`a di Milano-Bicocca\newline
and Istituto Nazionale di Fisica Nucleare, Sezione di Milano-Bicocca\newline
Piazza della Scienza, 3. I-20126, Milano, Italy}

\begin{abstract}
With neutrino oscillations now firmly established, 
neutrinoless double beta decay assumes great importance since it is one of the 
most powerful tools to set the neutrino mass absolute scale and establish whether the neutrino is a Majorana particle. 
After a summary of the neutrinoless double beta decay phenomenology, 
the present status of the experimental search for this rare decay is reported 
and the prospects for next generation experiments are reviewed.
\end{abstract}

\maketitle

\thispagestyle{fancy}

\section{Introduction}
The double beta decay is a second order weak transition which can be
energetically favored for some even-even nuclei belonging to $A$ even
multiplets.  
The $(A,Z)\rightarrow (A,Z+2) + 2e^- + 2\bar{\nu}_e$ double beta (\bb) decay process is 
allowed by the Standard Model and has been observed for many isotopes with 
lifetimes longer than $10^{19}$\,y. 
A more interesting process is the so-called neutrinoless double beta (\bbn) decay 
given by $(A,Z)\rightarrow (A,Z+2) + 2e^-$: this process violates
lepton number conservation and is therefore forbidden by the Standard Model.
The lifetime for the \bbn\ decay is expected to be longer than $10^{25}$\,y and
so far only
one evidence has been reported for $^{76}$Ge (see Sec.\,\ref{subsec:HK}).
For recent comprehensive reviews on this topic refer for example to Ref.\,\cite{reviews}.
\subsection{\bbn\ decay and neutrino physics}
Many mechanisms have been proposed for driving this decay, but the simplest one
is the ``mass mechanism'', where a light Majorana neutrino is exchanged.
Whatever is the mechanism actually causing  
the \bbn\ decay, its observation would imply that the neutrino is massive and is a 
Majorana particle (i.e. $\nu\equiv\bar{\nu}$).
For a light Majorana neutrino mediate \bbn\ decay, the rate is given by 
$[\tau_{1/2}^{0\nu} ]^{-1}= \langle m_\nu \rangle^2 F_N/ m_e^2$, where \me\ is electron mass and 
the nuclear structure factor $F_N$ contains the nuclear matrix element and the phase
space. The effective neutrino Majorana mass is given by 
$\langle m_\nu \rangle = | \sum_k m_{k} \eta_k |U_{ek}|^2|$, where $m_{k}$
are the mass eigenvalues of the three neutrino mass eigenstates $|\nu_k\rangle$,
$\eta_k$ are the CP Majorana phases ($\eta_k = \pm 1$ for CP conservation) and
$U_{ek}$ are the elements of the electron sector of the neutrino mixing matrix.
As suggested in \cite{mechanism} the exact mechanism causing the \bbn\ could be discerned by
measuring the decay rates for different isotopes.

With the help of the $\Delta m_{ik}^2=|m_{i}^2-m_{k}^2|$ and $\sin^22\theta_{ik}=f(|U_{ik}|^2)$ parameters 
determined by neutrino flavor oscillation experiments (see Ref.\,\cite{Neutrino2006} for the latest results), it is
possible to calculate \meff\ as a function of the unknown neutrino absolute mass scale and $\eta_k$ phases \cite{osc-bb}.
From this analysis two possible scenarios can be devised for the upcoming new generation 
experiments aiming at a 10\,meV sensitivity. 
(1) The \bbn\ decay is discovered with 
$\langle m_\nu \rangle \ge 10$\,meV: then the neutrino is a Majorana particle and the masses are either degenerate 
($m_{1}\approx m_{2}\approx m_{3}$) or follow an inverse hierarchy ($m_{3}\ll m_{1}\approx m_{2}$).  If neutrinos are
degenerate (for $\langle m_\nu \rangle \ge\approx 0.05$\,eV) then the absolute mass scale can be established.
(2) The \bbn\ decay is not observed and only an upper limit  
$\langle m_\nu \rangle \le 10$\,meV is set: then, if the neutrino is a Majorana particle, the  masses must
have a normal hierarchy ($m_{1} < m_{2}\ll m_{3}$).
\subsection{\bbn\ decay and nuclear physics}
To obtain \meff\ from the experimental observable \taubb\ the nuclear structure factor 
$F_N\equiv G^{0\nu}(Q_{\beta\beta},Z)|M^{0\nu}|^2$ must be
known. While the phase space $G^{0\nu}(Q_{\beta\beta},Z)$ can be precisely calculated, the nuclear matrix $|M^{0\nu}|$
contains the uncertain details of the nuclear part of the process. 
In fact there is a large spread in the nuclear matrix elements calculated by
different authors with different nuclear models \cite{nuclearmatrix} (see also references in Ref.\,\cite{reviews,
osc-bb}). 
Because of this spread also \meff\ is affected by large uncertainties (about a factor 3 on the average). 
Presently these uncertainties are a severe limitation to the potentialities of \bbn\ decay as a tool for neutrino
physics: it has been recently suggested \cite{rodin, kortelainen} that measured \bb\ decay lifetimes can be used to reduce the
spread in QRPA calculations. 
Nevertheless, it is important to search for \bbn\ decay of as many as 
possible candidate isotopes. 
\subsection{\bbn\ decay and CP-violation}
The observation of \bbn\ could be used to establish CP-violation associated with Majorana neutrinos in the lepton sector 
due to the $\eta_k$ Majorana phases. 
This issue has gained interest because it could provide an explanation for the observed 
baryon asymmetry in  the Universe through the leptogenesis theory \cite{leptogenesis}.
The Majorana phases could be constrained by simultaneous precise measurements of \meff\ and $\sum_i m_i$ (from cosmological observation) or 
$m_{lightest}$ (from $\beta$ end-point direct experiments). 
This possibility has been explored by many authors with opposite conclusions \cite{CP-yes, CP-no}. 
Indeed the task is experimentally  very challenging and maybe successful only for some values of the neutrino mixing matrix elements.
The possibility of success rely also on a strong reduction of the uncertainties in the mixing matrix elements, in 
\meff, in $\sum_i m_i$ (or $m_{lightest}$), and in the nuclear matrix elements.

\subsection{Experimental approaches to \bbn}
There are two approaches for direct  \bbn\ searches. 
In the first approach the \bbn\
active source is external to the detector: the experimental configuration usually
consists of foil shaped sources with two detectors (e.g. scintillators, TPCs, drift
chambers \dots) analyzing the electrons emerging from the foil.
Using tracking detectors a background rejection is possible studying the event topology. The limits of this approach are the 
energy resolution and the small source mass.

In the second approach the source is internal to the detector ({\it calorimeter}) and only the sum 
energy of the two electrons is measured. 
The signature for \bbn\ decay is therefore a peak at the transition energy \Qbb.
The detector can be a scintillator, a bolometer,  a semiconductor diode or a gas chamber.
Calorimeters can have large mass and high efficiency.
Depending on the technique, high energy resolution and also some tracking are possible.

From statistical considerations, the sensitivity $\Sigma(\tau_{1/2}^{0\nu})$ of a \bbn\ 
decay search is given by
$\Sigma(\tau_{1/2}^{0\nu}) \propto \epsilon\, i.a. (M
t_{\mathrm{M}}/(\Delta E\, bkg))^{1/2}$, 
where $\epsilon$, $i.a.$, $M$, $t_{\mathrm{M}}$, $\Delta E$ and $bkg$ are the detector efficiency,
the active isotope abundance, the source mass, the measuring time, the energy resolution and specific
background at \Qbb, respectively.
In case no background count is observed in the region of interest, the sensitivity becomes
$\Sigma(\tau_{1/2}^{0\nu}) \propto \epsilon\, i.a. M
t_{\mathrm{M}}$. In any experimental approach the various experimental parameters may be optimized up to some intrinsic
technical limit while working on the background level usually offers the best possibility of sensitivity improvement.

The background is therefore a fundamental issue in all \bbn\ searches: to reduce it, all passive (e.g. heavy shielding in underground
sites, material selection and purification) and active (e.g. Pulse Shape Discrimination, topology analysis 
through granularity and segmentation) measures must be taken.
However the background caused by the high energy tail of the continuous \bb\ spectrum cannot be
avoided and must be minimized by improving the energy resolution.

\section{Present and past experiments}
In the following a selection of the most sensitive experiments is presented (see Table\,\ref{table1} for a more
complete list of the best results to date for many isotopes).
\begin{table*}	
\caption{A selection of the past and present experiments giving the best result per
isotope to date. All given \taubb\ (\meff) are lower (upper) limits with the exception of the
Heidelberg-Moscow experiment where the 99.9973\% CL value is given. The spread in \meff\ is due to the uncertainties on
the nuclear factor $F_N$. 
\label{table1}}
\begin{tabular}{|rcccccccccc|}
\hline
\textbf{isotope} & \textbf{experiment} & \textbf{latest} & \textbf{$Q_{\beta\beta}$} &
\multicolumn{2}{c}{\textbf{i.\,a.}} & \textbf{exposure}           &
\textbf{technique}  &\textbf{material} & \textbf{\taubb}          & \textbf{\meff}  \\[0.5ex]%
{}      &{}          & \textbf{result} & \textbf{[keV]}            &\textbf{nat.}       & \textbf{enrich.} & \textbf{[kg$\times$y]}  &{}      &{}        & \textbf{[$10^{23}$\,y]} & \textbf{[eV]}     \\[0.5ex]
\hline
 \Ca & Elegant VI & 2004\cite{Elegant} & 4271 & 0.19 & -- & 4.2 & scintillator & CaF$_2$ & 0.14 & 7.2$\div$44.7 \\
 \Ge & Heidelberg/Moscow & 2004\cite{HM2004} & 2039 & 7.8 & 87& 71.7 & ionization & Ge & 120.0 & 0.44\\
 \Se & NEMO-3 & 2007\cite{NEMO3-latest} & 2995 & 9.2 & 97 & 1.8 & tracking & Se & 2.1 & 1.2$\div$3.2 \\
 \Mo & NEMO-3 & 2007\cite{NEMO3-latest} & 3034 & 9.6 & 95$\div$99 & 13.1 & tracking & Mo & 5.8 & 0.6$\div$2.40\\
 \Cd & Solotvina & 2003\cite{SolotvinaCd} & 2805 & 7.5 & 83 & 0.5 & scintillator & CdWO$_4$ & 1.7 & 1.7 \\
 \Te & Cuoricino & 2007\cite{Cuoricino-latest} & 2529 & 33.8 & -- & 11.8 & bolometer & \teo & 30.0 & 0.16$\div$0.84 \\
 \Xe & DAMA & 2002\cite{DAMA2002} & 2476 & 8.9 & 69 & 4.5 & scintillator & Xe & 12.0 & 1.10$\div$2.9 \\
 \Nd & Irvine TPC & 1997\cite{Irvine97} & 3367 & 5.6 & 91& 0.01 & tracking & Nd$_2$O$_3$ & 0.012 &3.0 \\
 \Gd & Solotvina & 2001\cite{SolotvinaGd} & 1791 & 21.8 & -- & 1.0 & scintillator & Gd$_2$SiO$_5$ & 0.013 & 26.0 \\
\hline
\end{tabular}
\end{table*}
\begin{table*}	
\caption{A selection of the proposed experiments. Except for CUORE and GSO all experiments use isotopically enriched material. 
Background {\it bkg} is calculated on an energy interval equal to $\sigma_E$. 
For all tracking experiments the quoted background is due only to the \bb\ tail. \label{table2}}
\begin{tabular}{|cccccccccccc|}
\hline
 \textbf{experiment} &  \textbf{isotope} &  \textbf{$Q_{\beta\beta}$} &  \textbf{tech.} &  \textbf{i.a.}  &  \textbf{mass} & 
 \textbf{$t_{\mathrm{M}}$}  &  \textbf{$\sigma_E$}  &  \textbf{{\it bkg}} &  \textbf{\taubb}  &   \textbf{\meff} &  \textbf{project}\\[0.5ex]%
{}      &{}          &  \textbf{[keV]} &              &  \textbf{[\%]}       &  \textbf{[kmol]}    &  \textbf{[y]}  &  \textbf{[keV]}     &  \textbf{[c/y]}       &  \textbf{[$10^{28}$\,y]} &  \textbf{[meV]}  & \textbf{status} \\[0.5ex]
\hline
 CANDLES IV+\cite{CANDLES} & \Ca & 4271 & scint. & 2 & 1.8& 5 & 73 & 0.35 &0.3 & 30 & R\&D (III: 5 mol) \\
 Majorana 120\cite{Majorana} & \Ge & 2039 & ion. & 86 & 1.6& 4.5 & 2 & 0.1 &0.07 & 90 & R\&D - reviewing \\
 GERDA II\cite{Gerda} & \Ge & 2039 & ion. & 86 & 0.5& 5 & 2 & 0.1 & 0.02 & 90$\div$290 & funded/R\&D (I: 0.3\,kmol) \\
 MOON III\cite{MOON} & \Mo & 3034 & track. & 85 & 8.5& 10 & 66&3.8 &0.17 & 15 & R\&D (I: {\it small}) \\
 CAMEO III\cite{CAMEO} & \Cd & 2805 & scint. & 83 & 2.7& 10 & 47 & 4 &0.1 & 20 & proposed \\
 CUORE\cite{CUOREprop} & \Te & 2529 & bol. & 33.8 & 1.7& 10 & 2 & 7.5 &0.07 & 11$\div$57 & construction \\
 EXO\cite{EXO} & \Xe & 2476 & track. & 65 & 60.0& 10 & 25 & 1&4.1 & 11$\div$15 & R\&D (1.5\,kmol) \\
 SuperNEMO\cite{SuperNEMO} & \Nd  & 3367 & track. & 90 & 0.7& -- & 57 & 10 &0.01 & 50 & R\&D \\
 DCBA-F\cite{DCBA} & \Nd & 3367 & track. & 80 & 2.7& -- & 85 & -- &0.01 & 20 & R\&D (T2: {\it small}) \\
 GSO\cite{SolotvinaGd} & \Gd & 22 & scint. & 22 & 2.5& 10 & 83 & 200 &0.02 & 65 & proposed \\
\hline
\end{tabular}
\end{table*}
\subsection{\Ge\ experiments and the evidence for \bbn\ decay}
\label{subsec:HK}
The {\bf Heidelberg-Moscow experiment}  (hereafter HM) searched the \bbn\ decay of \Ge\ using five High Purity Ge 
semiconductor detectors enriched to 87\% in \Ge. 
This experiment run in the Gran Sasso Underground Laboratory (Italy) from 1990 to 2003, totalling an exposure
of 71.7\,kg$\times$y (i.e. 820 moles$\times$y of \Ge). It is by far the longest running \bbn\ decay experiment
with the largest exposure. The experiment since the end of 1995 featured PSD on four crystals to reduce background by
separating Single Site Events (like \bbn\ decay events) from Multiple Site Events (like $\gamma$ interactions): the PSD
is applicable to 72\% of the full data set (i.e. 51.4\,kg$\times$y). The final background at \Qbb\ is about
0.11\,\ckky\ and it is attributed mainly to U and Th contaminations in the set-up materials.
The use of Ge detectors for a calorimetric \bbn\ decay search was first proposed in Ref.\,\cite{Fio1960} and the 
HM experiment is the best exploitation of this technique: it represents the Status-of-the-Art for the low background 
techniques and has been the reference for all last generation \bbn\ decay experiments.
After the conclusion of the experiment, part of the collaboration (hereafter KKDC) has reanalyzed the data \cite{HMNIM} claiming a
4$\sigma$ evidence for \Ge\ \bbn\ decay with a lifetime \taubb\ of about $1.2\times 10^{25}$\,y, corresponding
to a \meff\ of about 0.44 eV \cite{HM2004}. 
This claim has sparked a debate in the neutrino physics community \cite{reviews}
because the signal is indeed faint and close to other unexplained peaks.

{\bf Igex} \cite{Igex} is a similar experiment which run in Homestake (USA), Canfranc (Spain) and Baksan (Russia) from 1991
to 2000 with a total exposure of only 8.87\,kg$\times$y and a background at \Qbb\ of about 0.17\,\ckky:
its sensitivity is not enough to check the KKDC claim.
\subsection{Running \Se, \Mo\ and \Te\ experiments}
There are presently two  running experiments (Cuoricino and NEMO-3), 
which have chances to reach the sensitivity required to observe a \bbn\ signal at the level expected from the KKDC claim. 

{\bf Cuoricino} \cite{Cuoricino} is a calorimetric experiment using natural \teo\ cryogenic detectors to search for \Te\ \bbn\
decay. It runs in the Gran Sasso Underground Laboratory since 2003 and consists of 62 \teo\ crystals kept at a temperature of
about 10\,mK, arranged in a tower-like structure which is the base element of the future experiment 
CUORE (see Sec.\,\ref{subsec:CUORE}). The total \teo\ mass is about 41\,kg.
The total exposure to date is about 11.8\,\ky\ (i.e. 90.7\,\my\ of \Te) and the measured background at \Qbb\ is
about 0.18\,\ckky, mainly due to U and Th contaminations on the detector and surrounding copper surfaces.
This background level is obtained after applying anti-coincidence cuts between the detectors. The present 
90\% CL lower limit on \taubb\ is $3.0\times 10^{24}$\,y, corresponding to an upper limit on \meff\ of about
$0.16\div 0.84$\,eV \cite{Cuoricino-latest}.
With an exposure of about 120\,\ky\ (i.e. 3\,year running), Cuoricino would
reach a 1$\sigma$ sensitivity on \meff\ of about $0.1\div 0.6$\,eV.

{\bf NEMO-3} \cite{NEMO3} is a tracking detector experiment running in the Frejus Underground Laboratory (France). It uses a
drift chamber to analyze the electrons emitted by foils of different enriched materials. Interesting
\bbn\ decay sensitivities are expected only for the \Mo\ and \Se\ sources. The NEMO-3 detector can reject
the background by identifying $\gamma$s, e$^-$, e$^+$ and $\alpha$s. 
After measures taken in 2004 to suppress radon, the background is now about 0.5\,\cky\ (for \Mo): mostly composed by 
the \bb\ tail (60\%), \Tl\ in the foils (20\%) and radon (20\%).
The data analysis using a maximum likelihood applied to 3 kinematic variables gives a 90\%\,CL lower limit 
on \taubb\ of about $5.8\times 10^{23}$ ($2.1\times 10^{23}$)\,y for \Mo\ (\Se), corresponding to a limit on \meff\
of about $0.6\div 2.4$ ($1.2\div 3.2$)\,eV. 
In 2009, the expected 90\%\,CL sensitivity on \Mo\ \taubb\ is  $2\times 10^{24}$\,y,
corresponding to $0.3\div 1.3$\,eV for \meff. 
\section{Future experiments}
It is likely that presently running experiment will not be able to confirm or rule out the KKDC positive result:
therefore this will be the task for future experiments.
For a reliable confirmation, \bbn\ decay must be observed for different isotopes with similar \meff. 
The KKDC claim rejection requires a negative result from either a more sensitive \Ge\ experiment
or a much more sensitive experiment on a different isotope.
All proposed next generation experiments aim at sensitivities of about 0.01\,eV: whether the KKDC result is correct
or not, they will have good chances to observe \bbn\ decay.
The large sensitivity improvement (a factor 10 in \meff, i.e. a factor 10$^2$ on \taubb) must be obtained by scaling up 
to 1\,ton mass experiments and by further reducing the background. 
In order to perform high sensitivity searches for \bbn\ decay of as many different isotopes as possible,  
the isotope enrichment is becoming a hot topic: for many interesting isotopes large scale
enrichment is still both a technical and an economical problem. 
A strong effort is also demanded to nuclear theory to reduce the uncertainties in the nuclear matrix evaluation.

Table\,\ref{table2} gives some informations about the more well-defined projects.
Most of the projects presented here are at a very early R\&D stage, especially the ones in Sec.\,\ref{subsec:fut-scint} and 
\ref{subsec:fut-track}, and for all of them the predicted sensitivity heavily relies on the assumed background level. 
\subsection{Calorimetric experiments with ionization detectors}
The use of Ge ionization detectors is proposed for many future experiments because this is a well established experimental technique: 
it is relatively easy to scale up, it guarantees high energy resolution and it provides some background rejection by PSD and segmentation.
The main drawback is the high cost for the Ge enrichment and for the detectors themselves.
There is also the {\bf COBRA} proposal \cite{COBRA} for using CdZnTe diode detectors, but the technique, though promising, is still very young.

The Ge detector proposals follow two  opposite approaches, descending from the experience of the HM experiment and Igex. 
The first one attributes to the material
surrounding the detector the main responsibility for the background observed in the HM experiment, and therefore proposes to eliminate all 
this material by suspending bare Ge crystals in a highly purified cryogenic liquid. 
The second approach stems from the localization of the main source of the Igex background inside the Ge crystals due to cosmogenic activity.

The {\bf Majorana} experiment \cite{Majorana,Schoenert} belongs to the second group: 
the final aim is a 1\,ton experiment with segmented enriched Ge crystals in ultra low background cryostats.
The experiment will start in 2010 and a staged approach with 60\,kg detector assemblies is planned. For the first stage (Majorana 120)
two 60\,kg modules (114 detectors) will be installed either in the DUSEL or SNOlab underground laboratories.
Even minimizing cosmic ray exposure, the expected background of about 17\,\ckty\ (without cuts) is mainly
due to cosmogenics, since the activity in the surrounding materials would be avoided by careful material screening and selection: 
the application of PSD, granularity and segmentation cuts would further reduce the background 
to about 0.25\,\ckty, giving a 90\%\,CL sensitivity of $7\times 10^{26}$\,y in 5 years measuring time ($\langle m_{\nu}\rangle \leq 0.09$\,eV). 
The Majorana 120 phase would be able to probe the KKDC claim.
While a further enlargement to Majorana 180 is not yet settled, for a 1\,ton experiment a wider collaboration is foreseen and 
there is already a Memorandum of
Understanding  with the GERDA collaboration (see below).
The collaboration is presently going through a R\&D activity to study the segmented detectors (SEGA), to construct a prototype multi-crystal
cryostat (MEGA) and to understand and reduce the background.

The other approach is the one of Genius, GEM and GERDA.

The {\bf Genius} experiment \cite{Genius}, proposed by part of the HM experiment collaboration, consists of 1\,ton bare enriched Ge crystals suspended
in a 12\,m diameter liquid nitrogen tank. For a liquid purity of about $10^{-15}$\,g/g for U and Th, the expected background is about 0.2\,\ckty.
A 10\,year measurement would give a sensitivity of about $10^{28}$\,y ($\langle m_{\nu}\rangle \leq 0.015 \div 0.05$\,eV). This experiment could have
also an interesting sensitivity for real time solar neutrino detection and for cold Dark Matter. 
Although the authors believes that it is no longer worth to proceed with a 1\,ton experiment, 
given the positive result already claimed by KKDC, they set up the Genius Test Facility in the Gran
Sasso Underground Laboratory, where four 2.5\,kg Ge crystals have been run in liquid nitrogen \cite{Genius-TF}. 

Similar to Genius is the {\bf GEM} proposal \cite{GEM}: the main difference is the reduction of 
the amount of liquid nitrogen obtained by adding an external layer of pure water.

The new \Ge\ \bbn\ decay experiment in the Gran Sasso Underground Laboratory (also
known as {\bf GERDA}) \cite{Gerda, Schoenert} is similar to Genius and GEM  but has more compact dimensions. 
The driving idea is to scrutinize the KKDC evidence in a short time using the existing \Ge\ enriched detectors of 
the HM and Igex collaborations. 
The set-up consists of a liquid argon cryostat (4\,m diameter) immersed in pure water tank (10\,m diameter). 
The argon scintillation provides an additional active shielding, especially useful to reduce the effect of cosmogenic \Co\ in the detectors. 
The aim of the experiment Phase-I, planned to start in 2009, is to reduce the background to about 
0.01\,\ckky\  (mainly detector intrinsic) and to reach an exposure of about 15\,\ky\ using the 20\,kg of 
\Ge\ recovered from the HM experiment and Igex.
If the KKDC evidence is correct, GERDA would detect a 5$\sigma$ signal. 
In Phase-II, which is already funded, other 20\,kg of enriched and segmented Ge detectors will be added. A further reduction of
the background to 0.001\,\ckky\ and an exposure of 100\,\ky\ would give a  90\%\,CL sensitivity of about $2\times 10^{26}$\,y 
($\langle m_{\nu}\rangle \leq 0.09 \div 0.29$\,eV).
Presently the detectors for Phase-I are being refurbished by the manufacturer and the collaboration is working 
on the set-up (water tank, argon cryostat and infrastructures) to be installed in the Gran Sasso Underground Laboratory.
An R\&D activity is in progress to prepare and test the detectors for Phase-II. 

For a final 1\,ton \Ge\ experiment there are contacts between the GERDA and Majorana collaborations for a joint experiment using the 
best developed and tested technique \cite{Schoenert}.
\subsection{Calorimetric experiments with cryogenic detectors}
\label{subsec:CUORE}
The Cuoricino experiment has proved that also the cryogenic detection technique is mature for a next generation \bbn\ decay experiment. 
It is worth noting that almost all the interesting isotopes can be studied with cryogenic
detectors \cite{ScintMi,Chardin,MOON,CdMi}.
Cryogenic detectors have high energy resolution, can be scaled up to a 1\,ton size and their background can be reduced by segmentation.
Further background rejection can be achieved by hybrid detectors where, e.g., also scintillation or ionization are detected: 
these techniques can provide particle identification or position information \cite{ScintMi,Chardin}.
The drawbacks of this technique are the sensitivity to surface
contaminations, the difficulty to reduce close materials and the still cumbersome ancillary equipments required for cooling the detectors.

To date, the {\bf CUORE} (Cryogenic Underground Observatory for Rare Events) \cite{CUOREprop} is the only fully approved next generation 
1\,ton size \bbn\ decay  experiment: it is being built in the Gran Sasso Underground Laboratory where it is due to start data taking in 2011. 
CUORE will search for \bbn\ decay of \Te\ with a detector made of about 19 towers like the one of the running Cuoricino detector. 
988 natural \teo\ detectors will make up a 740\,kg granular and compact calorimeter containing 200\,kg of \Te. 
Even if it is possible to achieve a high sensitivity just with natural Te, the possibility of introducing enriched material in the core of 
the detector is still an open option for a second phase.
A background of about 1\,\ckty\ can be reached by exploiting the granularity and by reducing a factor 100 the surface contaminations 
observed in the Cuoricino experiment. 
This reduction is possible with the use of specially designed advanced cleaning processes which are being tested. 
Presently the cryogen-free dilution refrigerator and the infrastructure in the Gran Sasso Underground Laboratory are being built, while
the cryostat design is being completed. 
The copper for the construction has been procured and the crystal production is starting.   
With a background of about 1\,\ckty\ and an energy resolution FWHM of about 5\,keV, a 1$\sigma$ sensitivity 
on \taubb\ of about $6.5\times 10^{26}$\,y can
be reached in 5\,years ($\langle m_{\nu}\rangle \leq 0.011 \div 0.057$\,eV). CUORE is potentially also a good detector for cold Dark Matter and Solar
Axions \cite{CUOREpot}.
\subsection{Calorimetric experiments with scintillators}
\label{subsec:fut-scint}
Scintillators provide a relatively simple and well established instrument to search for \bbn\ decay of many interesting isotopes.
They can be extremely large and, in order to reduce the background, they can be immersed in the ultra pure liquids of
large solar neutrino experiments (e.g. Superkamiokande, SNO or Borex) using their photomultipliers. 
Their background can also be reduced by PSD.
The main drawback is the poor energy resolution which makes the \bb\ decay tail the main component of the background at \Qbb.
Moreover  photomultipliers and scintillators are often not enough radiopure for low background application. 

Most noticeable are the {\bf CAMEO} proposal \cite{CAMEO} to immerse CdWO$_4$ crystals in Borexino or CTF, the {\bf CANDLES} project 
\cite{CANDLES} to use CaF$_2$ crystals in a liquid scintillator active shielding to search for \bbn\ decay of \Ca, 
in spite of its exceedingly low natural isotopic abundance, and the possible use of
the Dark Matter self shielding 10\,ton liquid Xe {\bf XMASS} detector to look for \Xe\ \bbn\ decay \cite{XMASS1}. 
Even more challenging are the ideas to place liquefied Xe in high pressure transparent cells in
SNO \cite{XMASS2}, to dissolve Xe in Borexino \cite{caccianiga}, and to suspend scintillating
nanocrystals in SNO \cite{SNOLAB}.
\subsection{Tracking experiments}
\label{subsec:fut-track}
Tracking detectors could potentially avoid all background sources with the exception of the \bb\ decay tail. 
The main issue for this technique is therefore the  energy resolution.
In case a \bbn\ signal is detected, the reconstruction of the electron tracks would also provide a unique tool to distinguish the decay
mechanism from the electron angular correlation. 

The {\bf MOON} project \cite{MOON} consists of sandwiches of \Mo\ enriched foils, plastic scintillators, and scintillating fibers which  provide the 
energy and position measurements. MOON would be also a solar neutrino experiment. {\bf DCBA} project \cite{DCBA} proposes a Drift
Chamber Beta Ray Analyser with \Nd\ enriched foils.
There is also an Expression of Interest \cite{SuperNEMO} for a {\bf SuperNEMO} tracking detector with about 100\,kg of enriched isotopes (the
most interesting would be \Nd).

The {\bf EXO} project \cite{EXO} would deserve its own section because it is actually a calorimetric experiment with moderate tracking capability. It
is the evolution of the Gotthard experiment \cite{Gotthard} on \Xe, which used a high pressure Xe TPC. The EXO proposal adds the tagging of the
Ba atoms produced by the \Xe\ decay to completely suppress all backgrounds. A single Ba$^+$ ion would be detected by optical spectroscopy.
Presently there are still two open options for the detector: a high
pressure Xe TPC or a liquid Xe TPC where also scintillation would be detected to improve energy
resolution. The second option is the preferred one
because of its compactness (a 10\,ton detector would have a 3\,m$^3$ volume), but Ba tagging requires single Ba$^+$ ion extraction from liquid. 
Running for 5 years a 10\,ton Xe detector with an energy resolution of 1\% and with just
the \bb\ tail background, a sensitivity on \meff\ of about 11$\div$15\,meV could be reached. 
Presently a 200\,kg enriched liquid Xe TPC prototype without tagging is being installed at WIPP \cite{EXO}. 
Running for 2\,years the expected sensitivity
for EXO-200 is $6.4\times 10^{25}$\,y ($\langle m_{\nu}\rangle \leq 0.27 \div 0.38$\,eV).
\section{Conclusions}
\bbn\ experiments can establish whether the neutrino is a Majorana particle and fix the absolute neutrino mass scale.
Presently there is still only one experimental evidence for \bbn, which   has been claimed by the Heidelberg-Moscow experiment on
\Ge, but has not yet been confirmed nor ruled out.
Only the Cuoricino and NEMO-3 still running experiments have some chances to confirm this result.
Next generation experiments, aiming at a sensitivity on \meff\ of the order of 10\,meV, will settle the issue: presently only
the CUORE experiment is approved and being built full size. 
\end{document}